# Machine learning analysis of photometric data from the Dark Energy Survey

Elcio Abdalla,[1,2,3] Filipe B. Abdalla,[4,5★] Alessandro Marins,[4,5] Amilcar Queiroz,[6] Rafael M. Ribeiro[1] and Alex S. C. Souza[1]

[1] *Instituto de Física, Universidade de São Paulo, C.P. 66318, CEP 05315-970, São Paulo, Brazil*
[2] *Universidade do Estado da Paraíba, R. Baraúnas, 351, Universitário, Campina Grande – Paraíba, 58429-500, Brazil*
[3] *Centro de Ciências Exatas e da Natureza – CCEN, Universidade Federal da Paraíba, CEP 58059-970, João Pessoa – Paraíba, Brazil*
[4] *Department of Astronomy, University of Science and Technology of China, Hefei 230026, China*
[5] *School of Astronomy and Space Science, University of Science and Technology of China, Hefei 230026, China*
[6] *Unidade Acadêmica de Física, Universidade Federal de Campina Grande, R. Aprígio Veloso, Bodocongó, 58429-900 – Campina Grande, Paraiba, Brazil*

## ABSTRACT

In order to retrieve cosmological parameters from photometric surveys, we need to estimate the distribution of the photometric redshift in the sky with excellent accuracy. We use and apply three different machine learning methods to publicly available Dark Energy Survey data release 2 (DR2): (a) artificial neural network for photometric redshifts (ANNz2); (b) Gaussian processes for photometric redshifts (GPz); and (c) Keras, a deep learning application programming interface in PYTHON . We compare these different techniques applied to training data obtained from the Visible Multi-Object Spectrograph (VIMOS) Public Extragalactic Redshift Survey (VIPERS) survey. To deal with the incompleteness of the VIPERS catalogue, we use a space-partitioning data structure (`K-d Tree`) to estimate the reliability of the obtained photometric redshifts. We build a catalogue which is robust to the lack of training data in certain regions of colour space. We use the photometric data to create maps of overdensity as a function of the redshift distribution for more than 500 million galaxies. These maps split the sky into several onion-like redshift slices, which can be readily used for cosmological parameter estimation. On each angular slice, we create and present maps of the angular distribution of galaxies in that slice as well as an estimate of the redshift distribution, $n(z)$, related to the galaxy distribution of that slice, which is recovered from the redshift estimation methods. We achieve a subsample of DES galaxies, which are well matched to the VIPERS sample with an accuracy of the photometric redshifts with a $\sigma_{68} \sim 0.035$ and a catastrophic outlier rate of the order of 3 per cent.

## 1 INTRODUCTION

The two supernovae Ia (SNe) surveys published in the late 1990s (Riess et al. 1998; Perlmutter et al. 1999) show that the Universe has counter-intuitive properties – it is currently in an accelerated expansion phase. This unexpected property requires either an unusual energy-density content with a negative pressure or a modification of the theory of gravity itself.

The accelerated expansion is attributed to the so-called dark energy (DE), and the last two decades revealed a new era of cosmology in which the subject turned into a precise science with a wealth of data. The uncertainties in measurements of the cosmological parameters were reduced to less than a few per cent by recent experiments. The astronomical community has acquired access to a considerable amount of survey releases with increasingly better data sets, such as the Dark Energy Survey (DES; Abbott et al. 2021), the Panoramic Survey Telescope and Rapid Response System (Pan-STARRS; Chambers et al. 2016), the Deep Extragalactic Evolutionary Probe (DEEP; Newman et al. 2013). In the Infrared part of the spectrum, we have the *Wide-field Infrared Survey Explorer* (*WISE*; Wright et al. 2010) and the UltraVISTA survey (UltraVISTA; McCracken et al. 2012).

As a community, we have also made significant leaps in spectroscopic surveys with the Baryon Oscillation Spectroscopic Survey (BOSS; Dawson et al. 2012), the VIMOS Public Extragalactic Redshift Survey (VIPERS; Scodeggio et al. 2018a), the Dark Energy Spectroscopic Instrument (DESI; Flaugher & Bebek 2014), the Galaxy And Mass Assembly (GAMA; Driver 2008), and the Cosmic Evolution Survey (COSMOS; Weaver et al. 2023). In summary, these projects represent a significant increase in our knowledge about the Universe in large-scale structures.

The behaviour of DE is determined by the ratio $w$ of its pressure to its energy density (Dodelson 2003; Komatsu et al. 2009; Reid et al. 2010; Melia 2015; Poplawski 2019; de Boer et al. 2022). The $w$ parameter can either be a constant or vary with time (Astier et al. 2006; Riess et al. 2007; Wood-Vasey et al. 2007; Bamba et al. 2012). Probing the evolution of $w$ (by looking at its impact on distances as well as on the growth of fluctuations) will let us determine whether DE is a cosmological constant, possibly a fundamental scalar field (Wang et al. 2016; Abdalla & Marins 2020), an effect of modified gravity, or a number of many other possible alternatives. Cosmological measurements have so far failed to reveal any deeper physical picture.

★ E-mail: filipe.abdalla@gmail.com

In order to take advantage of each different cosmological survey, it is necessary to use the most pertinent information that is encoded into the data. In our case, the cosmological redshift, z, is the variable that needs the most careful calibration. In fact, high-precision redshift calibration of cosmological probes is one of the most important goals of cosmological missions (Tegmark et al. 2006; Blake et al. 2011; Bautista et al. 2020).

The measurement of the redshift z can be made in two ways: *spectroscopically* or *photometrically*. The advantage of using photometric redshift relies on them being cheaper and faster to measure, although less accurate than spectroscopic redshift. However, in the case where we have a very large amount of data, photometric techniques can be useful and statistically powerful. This is the case for observations such as those obtained from the DES,[1] Sloan Digital Sky Surveys (SDSS[2]), and the Pan-STARRS.[3]

As mentioned in (Rivera et al. 2018), the effectiveness of photometric redshift methods relying on a training set depends on whether the training set itself is a representative sample of the photometric data set. In this paper, we aim to successfully apply machine learning methods capable of predicting photometric redshift with excellent accuracy to a subset of the public release of the DES survey (DR2) and provide an alternative catalogue to that provided by the collaboration that can be used by the community. To do this, we trained machine learning algorithms to perform this task; we used software that works in PYTHON, called KERAS (Chollet et al. 2015) as well as the publically available ANNz2 (Sadeh, Abdalla & Lahav 2016) and GPz (Almosallam et al. 2016a).

Furthermore, we implemented an estimator which estimates how representative the training set galaxies are based on their photometry. By using a recursive partition structure in a 'multidimensions' system (`K-d Tree`; Bentley 1975; Gao, Zhang & Zhao 2008) to store the galaxies in groups, we calculated the probability that each of these galaxies is representative of the training set. Beyond that, we estimate that the photometric redshift has a small error in relation to its true value.

For this work, we use the VIPERS survey as the spectroscopic catalogue to provide the spectroscopic redshift and the photometric information of galaxies for our machine learning. We used the DES as the photometric data to calculate the photometric redshift of millions of galaxies.

The data explored here are clearly well explored within the DES collaboration, and while we recognize that collaboration work is imperative for an excellent analysis of this kind of large data set, we would like to highlight that there are big benefits of other analyses carried out outside collaboration work. First, it is important to have parallel analysis of many types of data sets for corroboration of results, for instance Hütsi (2006) has reanalized some of the original SDSS DR4 data for BAO studies and has corroborated the collaboration results; this in itself is of great scientific value. The same was done by reanalysis of Planck data, raising discussions on sources of systematic effects on the data (Spergel, Flauger & Hložek 2015). The value of open and independent re-analyses cannot be understated.

Furthermore, the analysis here has other scientific benefits that a simple reanalysis does not have. While there are several analyses of photometric redshifts within the DES collaboration, if we look solely at the catalogues provided in the data release, we only find results arising from the template-fitting photo-*z* BPZ (Benítez 2000) and the nearest-neighbour redshift estimates from DNF photo-*z* estimator, which were only available in previous releases, such as Y3 GOLD Sevilla-Noarbe et al. (2021), and did not include the DR2 sample at the time this work was conducted. However, the DES Collaboration has recently begun gradually making the Y6 GOLD catalogues publicly available. Additionally, some of the more complex photo-*z* analyses are still not included in the general DR2 catalogue release. Furthermore, these internal collaboration releases the collaboration has produced have inhomogeneous photo-z training sets for the entirety of the DES galaxies. This makes it difficult to assess the reliability of the results in a simple and confident way. The DES analysis do have a smaller set of galaxies which is much more homogeneous in their selection, namely a set of redmapper galaxies arising from cluster-finding algorithms (Rykoff et al. 2016), which have much better photo-*z*; however, these account for only a small fraction of the full catalogue (around a per cent). Our analysis presented here as we will see, accounts for the majority of the catalogue (of the order of half of the galaxies); although a good fraction is removed by our selection algorithm, we provide a middle of the road approach to the photo-*z* solutions given by the DES collaboration, one where the homogeneity of the sample is much greater than the full sample from DES but is much larger, though less accurate, than the redmapper sample and associated clusters of galaxies within from DES. This highlights the importance of other analysis that are independent from the standard collaboration analysis, such as Ferreira & Reis (2025).

The work is divided as follows: in Section 2, the data used in our work is described. In Section 3, we specify the three methods we use, as well as the data sets and the metrics used to compare them. The galaxy selection function is explained in Section 4, including the K-d Tree data structure, followed by results and discussion in Section 5. Finally, in Section 6, we summarize our main results and conclusions.

## 2 DATA

The process of taking spectroscopic redshift data is time-consuming, expensive (as it requires dedicated time from large facilities), and challenging to do for large numbers of galaxies. Conversely, photometric redshifts are much cheaper as the data required in order to obtain them involves shorter exposures, even if we consider the time needed for multiple filters.

In this work, we used the magnitudes of the galaxies in the $m_g$, $m_r$, $m_i$, $m_z$ and $m_Y$ filters from DES DR2, as described in (Sánchez et al. 2014), in our training process. Combining all of these filters, we have a wavelength range from 4000 Å to 11 000 Å. This complete photometry set yields the photometric redshift for the objects in the field, whereas spectroscopy usually needs to be done for a subset of galaxies (Massarotti et al. 2001; Mo, Van den Bosch & White 2010; Rivera et al. 2018; Davidzon et al. 2019).

There are two ways of computing the photometric redshift. The first is through template-fitting methods that use a set of standard galaxy spectral energy distributions (SED; Connolly et al. 1995; Bolzonella, Miralles & Pelló 2000); the second one is through the training-set method, which is a characterization for the redshift as a function of photometric parameters (Massarotti et al. 2001; Rivera et al. 2018).

[1] https://www.darkenergysurvey.org/
[2] https://www.sdss.org/
[3] https://panstarrs.stsci.edu/

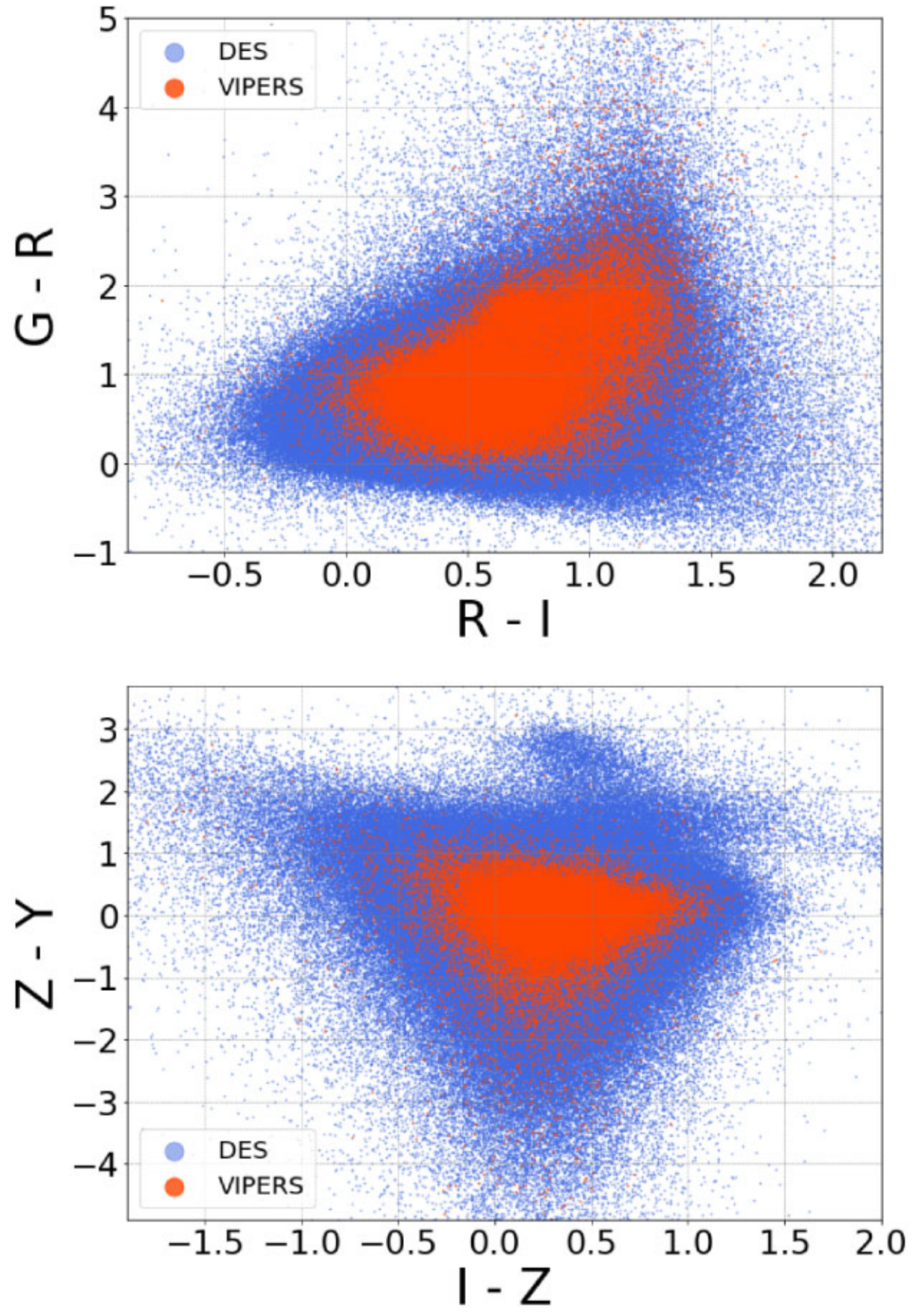

**Figure 1.** Colour–colour diagrams of the inputs used on this paper. In both images, the red dots represent the VIPERS galaxies, while the blue dots represent DES galaxies. It is possible to see that DES galaxies take a larger area of the plot than VIPERS galaxies. The first image is a scatter plot of the colours $m_r - m_i$ and $m_g - m_r$, and the second image is a scatter plot of the colours $m_z - m_Y$ and $m_i - m_z$.

The training-set methods have been described in several places in the literature; among such methods are polynomial fitting (Connolly et al. 1995), nearest neighbours (Budavári et al. 2001), neural networks (Collister & Lahav 2004a), Gaussian processes (GPs; Almosallam, Jarvis & Roberts 2016b), to name but a few. These methods are based on the use of a training set of galaxies to help configure hyperparameters of a machine learning technique. The data set determines a functional relationship between photometric parameters (e.g. magnitudes, colours, and morphological indicators) and redshift itself. In this paper, we decided to use the data from VIPERS as the training set, since it is well-matched (although not perfectly matched) to the photometric data set. We use such spectroscopic redshifts to evaluate the photometric redshift of the DES (Abbott et al. 2018, 2021).

The photometric data on both catalogues are given in the form of magnitude through the AB system. Fig. 1 is a plot representing the colour–colour scatter plots ($m_r - m_i \times m_g - m_r$, and $m_i - m_z \times m_z - m_Y$) used in this work. It shows the coverage of the DES and VIPERS in the colour–colour planes. The image contains about 46 thousand VIPERS galaxies (red dots) and 600 thousand DES galaxies (blue dots). We can see that there are clear regions where the VIPERS galaxies cover well the DES photometric sample and regions where this is not the case.

**Table 1.** The values used on the parameters of the ANNz2 code in both DR1 and DR2. Objects were matched with VIPERS objects based on their RA and DEC, and photometric redshifts were obtained based on their DERED magnitudes.

| | DES columns |
|---|---|
| Match | COADD_OBJECT_ID RA DEC HPIX 32,64,1024,4096,16384 |
| Machine learning | MAG_AUTO_G,R,I,Z,Y_DERED MAGERR_AUTO_G,R,I,Z,Y |

## 2.1 Dark Energy Survey

The DES (Abbott et al. 2018, 2021) is an optical and near-infrared public survey which has as a main objective to improve our knowledge of DE and cosmic acceleration by mapping galaxies, detecting SNe, and finding patterns of cosmic structure (To et al. 2021).

To achieve such objectives, four complementary observable data were considered by the DES collaboration at the time: type Ia SNe, the large-scale clustering of galaxies, galaxy clusters number counts, and weak gravitational lensing (The Dark Energy Survey Collaboration 2005). These probes of the DE are conducted using two distinct multiband imaging surveys: a 5000 $\deg^2$ in the *grizY* passbands and a 27 $\deg^2$ deep supernova survey observed in the *grizY* passbands (Kessler et al. (2015)).

This survey consists of images captured by a telescope located at the Cerro Tololo Inter-American Observatory (CTIO) during 345 nights in the course of 3 yr (Abbott et al. 2018). A new paper with the incorporation of more data collected over a period of 6 yr (including the 3 yr from DR1) was published in 2021 (Abbott et al. 2021). In order to retrieve the data used in this work, we used the publically available DESaccess,[4] which is an interface that allows the user to access the DES data.

The first release (DR1) of the DES catalogue has almost 400 million astronomical objects, of which around 80 per cent are galaxy and star samples (Abbott et al. 2018). Moreover, the second release (DR2) of the DES catalogue has almost 700 million astronomical objects with a similar fraction of galaxies and stars as the DR1 catalogue (Abbott et al. 2021). Each one of them has three tables with information on all the astronomical objects. We decided to use only the main table that contains the data necessary for our work. The columns used in our work are described in Table 1.

For this work, only photometric data was required, apart from spatial localization. These main tables contain the necessary information to estimate photometric redshifts and later derive cosmological information.

## 2.2 VIMOS public extragalactic redshift survey

The VIMOS Public Extragalactic Survey (VIPERS) is a spectroscopic survey performed at the European Southern Observatory (ESO) on the Very Large Telescope (VLT). Its main purpose is the measurement of a vast system and cosmological parameters at an epoch of half of the Universe's current age. The survey studies and characterizes the general galaxy population within a well-defined redshift range and is complemented by extensive photometric information. We use the public data release 2 (PDR-2; Scodeggio et al. 2018b), which contains redshifts, spectra, CFHTLS (Canada–France–Hawaii Telescope Legacy Survey) magnitudes, and ancillary information (as masks and weights) for a complete sample of more than seventy thousand galaxies. The data set from VIPERS (Scodeggio et al. 2018b) was collected from a Visible Multi-Object Spectrograph (VIMOS).

[4] https://des.ncsa.illinois.edu/releases/dr2/dr2-access/dr2-ncsa

The survey sample is magnitude limited to $m_i \leq 22.5$, with an additional selection by the colour that aims to homogenize it over the area that implies the removal of low-redshift galaxies ($z < 0.5$). The galaxies selected for the survey were those that obeyed $(m_r - m_i) > 0.5(m_u - m_g)$ or $(m_r - m_i) > 0.7$. We expect there to be differences between the colour cuts used by VIPERS, which come from CFHTLS and those available in the DES survey, although this difference would be expected to be small as the filters are very similar.

The VIPERS survey extended the volume explored by its predecessors, such as the VIMOS VLT Deep Survey (VVDS; Le Fèvre et al. 2013; Scodeggio et al. 2018b) and the ZCosmo (Lilly et al. 2009). Furthermore, it has double the sampling of the VIMOS spectrograph over the range $0.5 < z < 1.2$, essentially extending the exploration to a much larger volume of the Universe and with a high density of galaxies (Guzzo et al. 2014). The median redshift of the VIPERS survey is $z \simeq 0.7$.

When the sample is magnitude limited, we can have up to 30 per cent of stellar contamination in a galaxy survey (Garilli et al. 2008). In order to distinguish galaxies and stars, it was necessary to classify each object by its morphological proprieties. The VIPERS team used photometric data from CFHTLS combined with spectroscopic information from the VVDS-DEEP survey and the VVDS-Wide survey. The VVDS-DEEP catalogue contains more than ten thousand astronomical objects, such as galaxies, AGNs and stars with $m_i < 24$, while the VVDS-Wide catalogue contains more than eleven thousand galaxies and seven thousand stars with $m_i \leq 22.5$ (Le Fèvre et al. 2005). Since both catalogues are magnitude-limited, they are ideal for testing the completeness and contamination of the galaxy selection. The method to classify the VIPERS objects combines both its measured size object (half-flux radius $r_h$[5]) and SED.

The astronomical objects were divided into two groups according to their magnitude in filter i by the VIPERS team. The first group contained objects with $17.5 < m_i < 21$ and the second group with $m_i \geq 21$. For the brightest group, the half-flux radius was used to classify the objects, while for the other group, the two techniques ($r_h$ and SED fitting) combined were used.

For the first group, a Gaussian distribution with the value of the half-flux radius was made. From this distribution, VIPERS calculated its mean $\mu_{r_h}$ and its standard deviation $\sigma_{r_h}$ (Guzzo et al. 2014). As a consequence, the stars obey the equation:

$$r_h < \mu_{(r_h)} + 3\sigma_{(r_h)} \quad . \tag{1}$$

For the second group, small galaxies could be mistaken for stars through the use of $r_h$. In order to avoid this, the VIPERS team used the SED template fitting for the five filter bands along with a photometric redshift code called `Le Phare` (Ilbert et al. 2006; Arnouts & Ilbert 2011). This code criterion was: if the objects obey $r_h < \mu_{(r_h)} + 3\sigma_{(r_h)}$ and $\log_{10}(\chi^2_{star}) < \log_{10}(\chi^2_{gal}) + 1$, they are classified as stars and discarded from the sample.

This selection removed 21 per cent of the catalogued objects because they were classified as stars and 32 per cent of galaxies because they had low redshift. The remaining 47 per cent objects were released as part of VIPERS. This final data set was used in the present work.

To begin our work, we cross-match galaxies from the VIPERS data set with those in the DES data set. The cross-matching is done in the Right Ascension (RA) and Declination (DEC) of galaxies on both surveys; if the difference in spatial distance was less than 1 arcsec, they were considered to be a match.

## 3 ESTIMATION METHODS

Photometric redshift estimation is a technique that depends on the photometric properties of the galaxy population. This work aims to produce photometric redshifts for DES with three different estimation methods: ANNz2[6] (Sadeh et al. 2016), GPz[7] (Almosallam et al. 2016b), and KeraZ, the Keras (Chollet et al. 2015) applied to photometric estimation.[8] The first two methods are public algorithms, while the last one is a library-based model, which we have adopted for this work.

For all methods, we separate the data set into a training set, a test set, and a **validation** set. The percentage of the data that went to the training set, validation set, and test set is 65 per cent, 10 per cent, and 25 per cent, respectively. The division was made to avoid overfitting. We evaluate the performance of a method for given parameters on a test set, which is not used in the training and validation process.

Each of the learning algorithms uses the training set as well as the validation sets to check which are the optimal hyperparameter values that will generate a good predictive model (Xu & Goodacre 2018). The test set is used to confirm the predictions made by the learning algorithm. In the following subsections, we will briefly describe each training model mentioned and used in this work.

### 3.1 ANNz2

ANNz2 (Sadeh et al. 2016) is an updated version of the public photometric redshift estimation software developed by Collister & Lahav (2004b) (ANNz). The algorithm used in ANNz2 is implemented within the ROOT C++ software framework,[9] which has several machine learning models allowing the user to have greater control, as it also features a Python interface. The package used in this work is the Multivariate Data Analysis (TMVA; Hoecker, Speckmayer & Stelzer 2007), which employs artificial neural networks and improved regression trees.

An artificial neural network consists of an input layer, one or more hidden layers, and an output layer with the expected result. In our case, the structure used is called a multilayer perceptron (Menzies et al. 2015; Abirami & Chitra 2020). In such a structure, the first layer, the input, is the colour or magnitude of each galaxy, which is then connected with the following ones. Each layer has several neurons with weights defined by the structure of the network. Each intermediate layer provides a collection of response functions, which are represented by various activation functions such as sigmoid or tanh functions. The last layer, called the output layer, displays the desired photometric redshift or, alternatively, a probability distribution function (PDF) of the redshift.

[5]The half-flux radius is the radius of the circle around the astronomical object that contains half of its flux or energy.

[6]https://github.com/IftachSadeh/ANNZ
[7]https://github.com/OxfordML/GPz
[8]https://keras.io/
[9]https://https://root.cern/

**Table 2.** The values used on the parameters of the ANNz2 code in both DR1 and DR2.

| Parameter | Definition | Value | |
|---|---|---|---|
| | | DR1 | DR2 |
| nMLMs | Number of MLMs | 100 | 100 |
| minValZ | Min. value for redshift | 0.0 | 0.0 |
| maxValZ | Max. value for redshift | 2.7 | 3.5 |
| nErrKNN | Near-neighbours for error | 100 | 100 |
| ndOptTypes | MLM types | ANN BDT | ANN BDT |
| nPDFbins | Number of PDF bins | 200 | 200 |

These networks have a propagation algorithm that uses the ANN error function to evaluate the error in the output against the predicted result in the training data set. The error function chosen for this work was the mean-squared error (MSE). Given that the overfitting affects the effectiveness of the method's performance, we use part of the TVMA package, which includes regularization methods to avoid that. Consider the specific case of the *Bayesian Regularization*, the method adds a term to the ANN's error function, equivalent to the negative value of the log-likelihood of the training data, given the network model.

ANNz2 also uses boosted regression trees, which are binary trees combined with boosting method (Schapire 1990) that separate the data set into groups according to the constraint defined in the parameters. Each galaxy is classified in a leaf in the tree, represented by a sequence of cuts that select a specific combination of input values of the target. The training decision tree defines the splitting criteria for each node. At each node, the division is determined at a cut-off median value of the output that best divides the training set. The training stops when the minimum number of training objects in a single group or cell is reached, obeying a threshold value that can vary from 0.1 per cent to 1 per cent of the number of training objects per node. The boosting process involves training multiple classifiers using the same data sample, where the data is re-weighted differently for each tree. A combined estimator is then obtained from the weighted average of the trees in this boosting process (as in Sadeh et al. 2016).

The solutions are attributed uncertainty values calculated with the K-nearest neighbour method (see Oyaizu et al. 2008). The values of entries and parameters used in this work are in the Table 2.

### 3.2 GPz

The Gaussian processes for photometric redshifts (GPz)[10] (Almosallam et al. 2016b) method used in this work is a Bayesian machine learning code written in MATLAB and PYTHON. GPz is a supervised non-linear regression algorithm used to make inferences about the redshift of a galaxy.

The method is outlined in Almosallam et al. (2015), and we summarize it here; we follow the same notation as in this paper. We define an input set

$$X = \{x_i\}_{i=1}^{n} \in \mathbb{R}^{n\times d} \quad , \tag{2}$$

where $n$ is the number of samples in the data set, and $d$ is the dimensionality of the set of inputs. In our notation, $x_i$ is a vector of dimension $d$. In our case, $d$ spans either the colours plus one magnitude or the magnitudes of the galaxies. The set of outputs is given by

$$y = \{y_i\}_{i=1}^{n} \in \mathbb{R}^{n} \quad , \tag{3}$$

which, in our case, are the redshifts of our objects.

[10] https://github.com/OxfordML/GPz

**Table 3.** The values used on the parameters of the GPz code in both DR1 and DR2.

| Parameter | Definition | Value |
|---|---|---|
| Method | GP method | VC |
| m | Number of basis function models | 25 |
| Heteroscedastic | Heteroscedastic noise | True |
| csl_method | Cost-sensitive | Normal |
| maxIter | Number of maximum iterations | 150 |
| maxAttempts | Max. iterations to attempt | 50 |

The observed target $y_i$ is generated by a set of non-linear functions $\phi$ of the input $x_i$ and an additive noise $\epsilon$. This can be written as

$$\phi(x_i) = [\phi_1(x_i), ..., \phi_m(x_i)] \in \mathbb{R}^{m} \quad , \tag{4}$$

where $y_i = \phi(x_i)w + \epsilon_i$, where $w$ is a vector of length $m$ of real-valued coefficients, or the parameters of the model following the definition of Almosallam et al. (2015) and the error (or noise) $\epsilon$ is taken to be normally distributed with zero average and variance $\sigma_n^2$, i.e. $\epsilon \sim \mathcal{N}(0, \sigma_n^2)$.

In Almosallam et al. (2016b), the radial basis function kernel is defined as

$$\phi_j(x_i) = \exp\left(-\frac{1}{2}(x_i - p_j)^T \Gamma_j^T \Gamma_j (x_i - p_j)\right) \quad , \tag{5}$$

where $P = \{p_j\}_{j=1}^{m} \in \mathbb{R}^{m\times d}$ is the set of basis vectors associated with the basis functions. Following the same definition as $x_i$, the vector $p_j$ also has a length of dimension $d$, having same dimensionality as $x_i$. However, the index $i$ represents the galaxy or training point and the index $j$ represents the index of the basis of dimension $m$.

The term $\Gamma_j^T\Gamma_j$ relates to the variable covariance, and the Gaussian process (GP) algorithm allows for different types of covariance matrices to be used. We performed tests and chose a GP with variable covariances (GPVC) that presented the best results. It is assumed that $\Gamma_j$ is unique for each basis function $j$.

In Table 3, we present the parameter values used as input for the GPZ code in this work. The GP goal is to maximize the probability of the observing target $y$ given some input $X$. The free parameters in the marginal likelihood which can be optimized are the kernel and the noise variance.

In classical GP models, the noise variance is defined as constant, while in GPz, the variance is used as an input-dependent function, which is determined together with the mean function. This feature allows the model to identify regions of the input space (for this work, the photometric space) where more data is needed in contrast to places where additional precision of information is needed. For further details, see Almosallam et al. (2016b).

Several combinations were tested using the test set results to decide which covariance to use. The configuration with the fewest errors was chosen.

Our goal for the best result is to have the minimum error possible and the code shows us, at each iteration, whether it has found the predicted redshift with the lowest RMSE. The code converged relatively quickly after 150 iterations in the conditions we ran.

The GPz model has the ability to evaluate the noise in the processing of estimating the redshift to recognize the different variances. If this option was disabled, the model would assume a global Gaussian noise variance. This speeds up the training process but would not give us much precision of the photometric redshift. We decided to enable (True) the Heteroscedastic noise.

### 3.3 KeraZ

Keras is an open-source software written in PYTHON in a deep learning application programming interface (API) which allows the user to write codes with a neural network quickly and simply (Chollet 2017; Vidal, nas & Álvaro Dósil Suárez 2021). It manages multiple machine learning languages and also allows the user to use different types of machine learning, such as classification, regression, time series, computer vision, and natural language processing, all in the same framework.

Keras implements layers, activation functions, and optimizers, which are used in general neural network theory. The network architecture can be either sequential (the code reads the layers linearly and in the order defined by the user; it is the most common type of neural network) or functional API (the code reads the layers in an arbitrary order defined by the user). In this work, we used the sequential model, which means that the output of each layer is the input of the following layer until the end of the model (Abadi et al. 2016; Seide & Agarwal 2016; Vidal et al. 2021).

On each layer, we define an *activation* feature. This feature defines the relationship between the inputs and the outputs. There are several pre-defined activation features in Keras that allow the user to create custom activation features.

We use a neural network with one hidden layer that contains 30 nodes. In this layer, we applied the tanh activation and a *batch normalization*[11] on the output. In addition to batch normalization, we also use regularizes to increase our chance of not getting overfit values (Chollet 2017). These techniques are essential for estimating values in any regression problem. We note that this is very similar to the original implementation of ANNz; however, it has the added flexibility of other activation methods and self-imposed regularization methods described above. For this reason, the addition of a constant value is discarded. The regularizer allows the layer's parameters or the activation function to be penalized during the optimization process.

To optimize our neural network, we choose the adaptative momentum estimation (Adam). In this algorithm, the learning rate[12] is computed individually for different parameters. This method is computationally efficient, has little memory requirement, and is well suited for problems with a large set of data (Kingma & Ba 2015; Reddi, Kale & Kumar 2018).

After choosing the optimizer, we can proceed with training the model. Thus, we followed the training steps of a machine learning model and divided our data set into training, validation, and testing. During the training, we used the architecture mentioned above, including all the features and the optimizer. We also added an early stop condition in which we check the loss function for 10 epochs, and if there is no change, we stop training. Finally, the estimated values are obtained by the output layer in which we use a linear activation, and the result is the single-solution photometric redshift.

It is worth noting that although Keras is not strictly speaking a machine learning technique, it does however have its internal procedures and methods for dealing with convergence of networks; it is therefore a specific implementation of an artificial neural network method. We can also underline that ANNz and ANNz2 are also not strictly speaking photo-$z$ methods, although they are currently accepted as such, given their wide use and their specific implementation and choices for the specifics for the neural network in such package. Given that Keras contains a set of features, including minimization and network activations, cost functions, and validation methods, establishing a robust framework that yields different results; in order to characterize these nuances, we will hereafter name our use of Keras for photo-z as KeraZ.

[11]Batch normalization is a node layer that centralizes and rescales the data around zero and normalizes it by the standard deviation

[12]It is an amount of change to the model during each step of the training.

### 3.4 Metrics

Here we compare the result of the three different methods (ANNz, GPz, and KeraZ) in evaluating the photometric redshift, we decided to use the metrics defined by Rivera et al. (2018).

The first metric is the $Bias$ that measures the normalized mean difference between the estimated photometric redshift and the spectroscopic redshift,

$$Bias = \left\langle \frac{z_{\rm phot} - z_{\rm spec}}{1 + z_{\rm spec}} \right\rangle . \tag{6}$$

It measures the systematic error that occurs within the model of machine learning due to incorrect assumptions in the process. The next metric is the scatter, $\sigma$, between the photometric redshift and the spectroscopic redshift,

$$\sigma = \left\langle \left( \frac{z_{\rm phot} - z_{\rm spec}}{1 + z_{\rm spec}} \right)^2 \right\rangle^{\frac{1}{2}} . \tag{7}$$

It measures how close the results of the photometric redshift are from the true redshift. The $\sigma_{68}$ estimator defines the error which brackets 68 per cent of a certain data set $U$ that have the smallest value of the term $|z_{\rm phot} - z_{\rm spec}|/(1 + z_{\rm spec})$,

$$\sigma_{68} = \max_{i \in U} \left\{ \left| \frac{z^i_{\rm phot} - z^i_{\rm spec}}{1 + z^i_{\rm spec}} \right| \right\} , \tag{8}$$

where $U$ is a subset of the full validation set in that redshift bin. This metric collects the 68 per cent galaxies that have the smallest value of the error.

The last metric evaluates the catastrophic outlier rate, which uses the number of galaxies whose photometric redshift is within the value defined by the outlier threshold $e$ to find the percentage of photometric redshifts that are considered to be a good result,

$$\mathrm{FR}_e = \frac{100}{n} \left\{ i : \left| \frac{z^i_{\rm phot} - z^i_{\rm spec}}{1 + z^i_{\rm spec}} \right| < e \right\} . \tag{9}$$

The term inside the braces is the number of galaxies that have $|z_{\rm phot} - z_{\rm spec}|/(1 + z_{\rm spec}) < e$ with $e = 0.15$ and $n$ is the total number of galaxies in our data set. It measures the percentage of galaxies whose photometric redshift values are within the margin of error.

## 4 THE DES/VIPERS GALAXY SELECTION FUNCTION

This analysis aims to make a reliable estimate of the photometric redshifts of a set of DES galaxies using a training set of VIPERS galaxies. However, the VIPERS training set is not fully representative of the full DES data set. Therefore, we need to perform a selection function analysis where we will obtain a likeness of any DES galaxy to an equivalent galaxy in the VIPERS data set. This will allow us to have a final catalogue where we can select the galaxies within the DES survey that have reliable photometric redshifts from VIPERS and those which are not covered by the VIPERS training set and, therefore, will not have reliable redshifts.

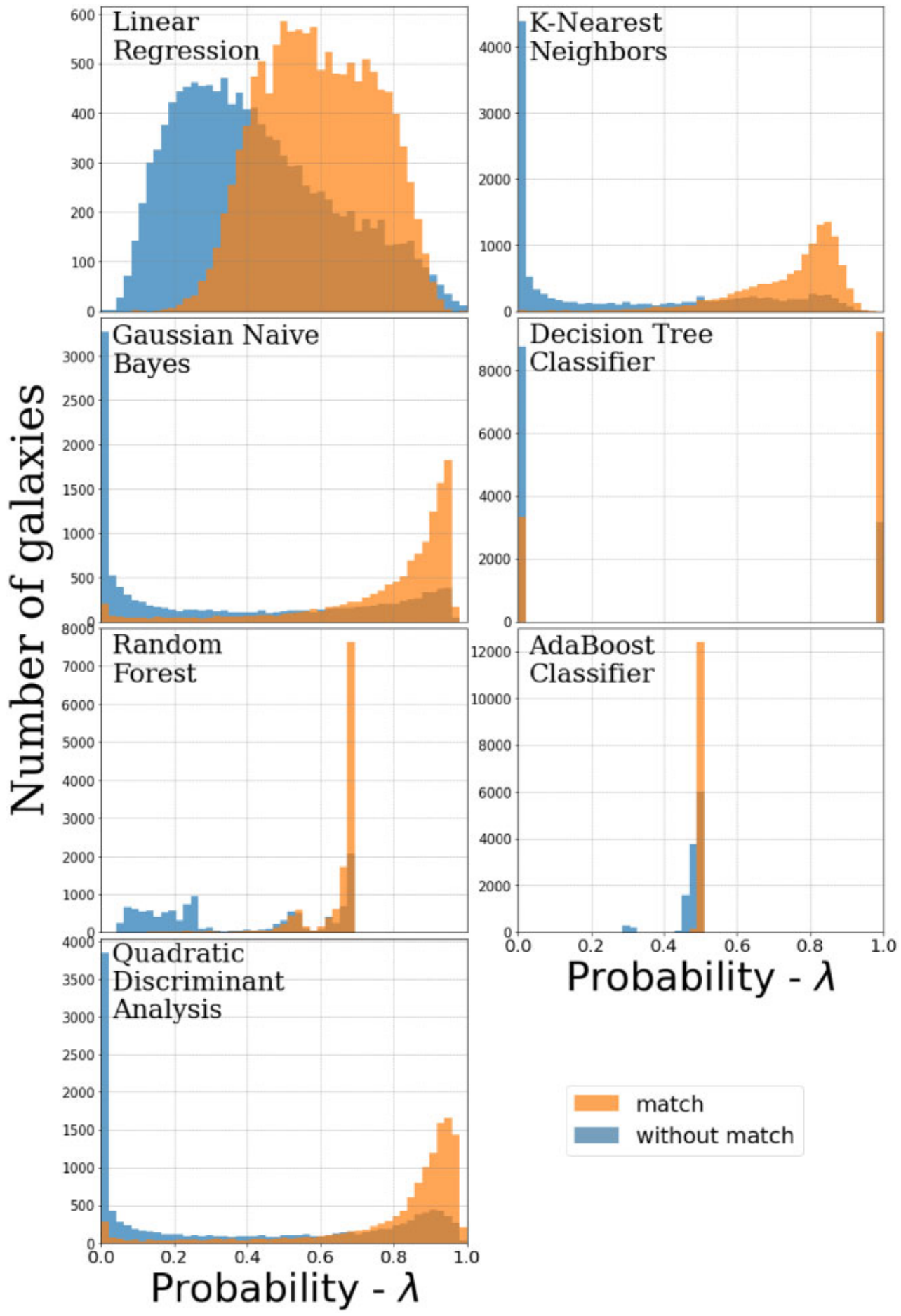

**Figure 2.** Histograms of the predicted probabilities of each DES galaxy being similar to the VIPERS galaxy applied to the test set. Each of the graphs was made using only one type of machine learning. The test set allows us to be able to predict how well we can separate the galaxies in both types, those not within the VIPERS data set (blue) and those within the VIPERS survey (orange).

The galaxy selection function we aim at producing refers to the process of selecting the subset of galaxies from the overall population of DES galaxies based on their likeness to the colour properties of the VIPERS galaxies. The goal of galaxy selection is to obtain a sample of the DES galaxies that is representative of the VIPERS population.

It is important to take the selection function into account when analysing the data from the DES DR1 and DR2 surveys, as the selection criteria can introduce biases or affect the statistical properties of the resulting sample. The selection function can also be used to estimate the completeness and efficiency of the survey, which will be important for interpreting the results and making statistical inferences about the overall population of galaxies in the survey. This will take into account regions of the survey which where we have low completeness, given that the VIPERS survey will not be able to identify galaxies at a high redshift (when, for example, the O II line falls beyond the wavelength range of the detectors); another example would be in regions of parameter space which have been selected out in colour space by the VIPERS team.

The DES DR2 survey (Abbott et al. 2018, 2021) covers an area of 5000 square degrees in the southern sky, while VIPERS observed a region of about 24 square degrees. Data from DES contain about 300 million galaxies, while those from VIPERS have about 90 thousand.

VIPERS galaxies have been selected with a specific set of colour cuts in order to select galaxies at high redshift (higher than 0.5), including a colour cut involving $u$-band data, which is not available by DES. For the purpose of attempting to recreate this selection function without the $u$-band data, we deploy machine learning methods that estimate whether the DES galaxies have similarities in colours with VIPERS galaxies. For every DES galaxy, we estimate a parameter $\lambda$, which is defined as a variable that ranges from 0 to 1. When it is closer to 1, the bigger our confidence in the photometric properties of that DES galaxy has galaxies in the VIPERS survey with similar colours. On the other hand, $\lambda$ close to 0 means that the galaxy from the DES survey is not got photometrically similar galaxies in the vicinity of the colour space to any galaxy from VIPERS reference sample, which implies that any redshift estimation that we do will have lower confidence.

The variable $\lambda$ can be interpreted as a probability of the given galaxy from DES being representative of the VIPERS sample.

### 4.1 Machine learning prediction of the selection function of DES galaxies

To determine the value of lambda, we use, as a tool, the distances between the galaxies in four-dimensional space of colours ($m_g - m_r$, $m_r - m_i$, $m_i - m_z$, and $m_z - m_Y$) as well as in five-dimensional space of colours and magnitude ($m_g - m_r$, $m_r - m_i$, $m_i - m_z$, $m_z - m_Y$ and $m_i$). We perform the analysis in both four- and five-dimensional spaces described above. The points representing galaxies are used to compare the distances between a given galaxy and those galaxies contained within the training set (65 per cent from the original data set). These distances are used to train the machine learning methods to classify whether the DES galaxy can be considered photometrically similar to the VIPERS galaxies.

We have compared results with different distance estimates, such as Euclidean, Manhattan, Minkowski, Chebyshev and Mahalanobis distances. For this work, we chose the one it had the best rate of successful classifications for the testing galaxies, it was obtained with the Minkowski distance, defined as

$$\mathrm{dist}_{(\mathrm{Minkowski})} = \left( \sum_{j=1}^{K} |x_j - y_j|^p \right)^{\frac{1}{p}}, \tag{10}$$

where $x_j$ represents the colour/magnitude $m_j$ from galaxy $x$ and $y_j$ is the colour/magnitude $m_j$ from galaxy $y$. $p$ is a constant that we choose to be equal to 2, and $K$ is the number of colours/magnitudes. Since we have run our methods in both colour and magnitude spaces, we have therefore used two K-d Trees with different dimensions, $K = 4$ and $K = 5$. Each one of the magnitudes used ($m_g$, $m_r$, $m_i$, $m_z$, $m_Y$) has an uncertainty in magnitude estimation which was taken into account.

We used a total of 7 classifier methods for these data sets. In all cases, the estimator of $\lambda$ is an interpolated value for the objects in the training and validation sets which belong to the VIPERS data set within the DES galaxies, where the galaxies that are within VIPERS are given a nominal value of 1 and those which are not being given a nominal value of 0. The first method presented, a Linear Regression, uses a linear interpolation in K-dimensions (this time with $K = 4$) with the lowest quadratic error. The second method used is a Random Forest Regressor method that consists of using a decision tree that uses the data set to evaluate the probability of galaxies being photometrically similar or not. This training method did not give us a particularly good result since it was unable to differentiate the galaxies with match properly and the ones without match, as we can see in Fig. 2 that the blue histogram (namely, *without match*) and the orange histogram (namely, *match*) overlap significantly. Other methods used were a K-nearest neighbours classifier, a decision tree classifier, random forests within the ANNz2 package, a Bayesian method, an Ada Boost Classifier and a quadratic Discriminant analysis. The difference between the 'Decision Tree' and 'Random Forest' methods is that the 'Random Forest' uses the basis of the 'Decision tree' method generates n-Decision trees and calculates n-results for each tree that ultimately, will determine the classification of the galaxy. That is, one method uses only one 'tree' (Decision Tree) the other uses 'n-trees' (Random Forest).

In Fig. 2, we present histograms for the estimated values of lambda for the testing set in each of the machine learning methods used in this work. All methods look at the training set composed of both DES and VIPERS galaxies, where the DES galaxies are assigned a value of 0 and the VIPERS galaxies have an assigned value of 1. We are able to make, therefore, two histograms, one for galaxies with DES with a match to VIPERS and another for those without. The machine learning method then chooses the most appropriate value (between 0 and 1) for the validation galaxies via training, and the plot shows the evaluation of the method for the testing galaxies.

Since each galaxy has a designated $\lambda$ from the training method, this implies that if the galaxy has a bigger $\lambda$, then we would consider it more similar to a VIPERS galaxy; if not, we would consider it as not similar.

In order to evaluate the performance of these methods (Fig. 3), we used the receiver operating characteristic curve, henceforth called the ROC curve (Fawcett 2006), which is a graphical way of characterizing the efficiency of a binary classifier system. This graph is made with the true positive rate (Sensitivity) as a function of the false positive rate (fall-out). The quality of each binary classification method is given by the area below the line drawn by each method.

For this evaluation, we divided the data set into a training group (about 30 per cent of the total data) and a test group. For the training group, we fitted the machine learning method such that the estimator could study the data set, and we used the test group to predict $\lambda$. The test group was a control group by comparing it with the $\lambda$ predicted since, for each galaxy, we had the real data (true values) and the predicted data. From this, we were able to separate the predicted data into four different groups: True Positive is the data which $\lambda$ of the galaxy is close to 1, and the real data is 1; False Positive is the data which $\lambda$ of the galaxy is close to 1 and the real data is 0; False Negative is the data which $\lambda$ of the galaxy is close to 0 and the real data is 1; True Negative is the data which $\lambda$ of the galaxy is close to 0 and the real data is 0. Using these divisions, we were able to plot the ROC curve of the sensitivity as a function of the fall-out, or alternatively speaking, as a function of $\lambda$, which are given by

$$\mathrm{TPR} = \frac{\text{True Positive}}{\text{True Positive} + \text{False Positive}}, \tag{11}$$

$$\mathrm{FPR} = \frac{\text{False Positive}}{\text{True Negative} + \text{False Negative}}. \tag{12}$$

Organizing all these results, we decided to use the Stacking Regressor from the *Sklearn* package, which combines different machine learning methods in order to obtain the best ROC curve possible. We can see that the best classifier in our case is the k-nearest neighbours method, which effectively looks at the k-nearest neighbours in this four- or five-dimensional space and sees which fraction of them is formed of galaxies which belong to the VIPERS survey.

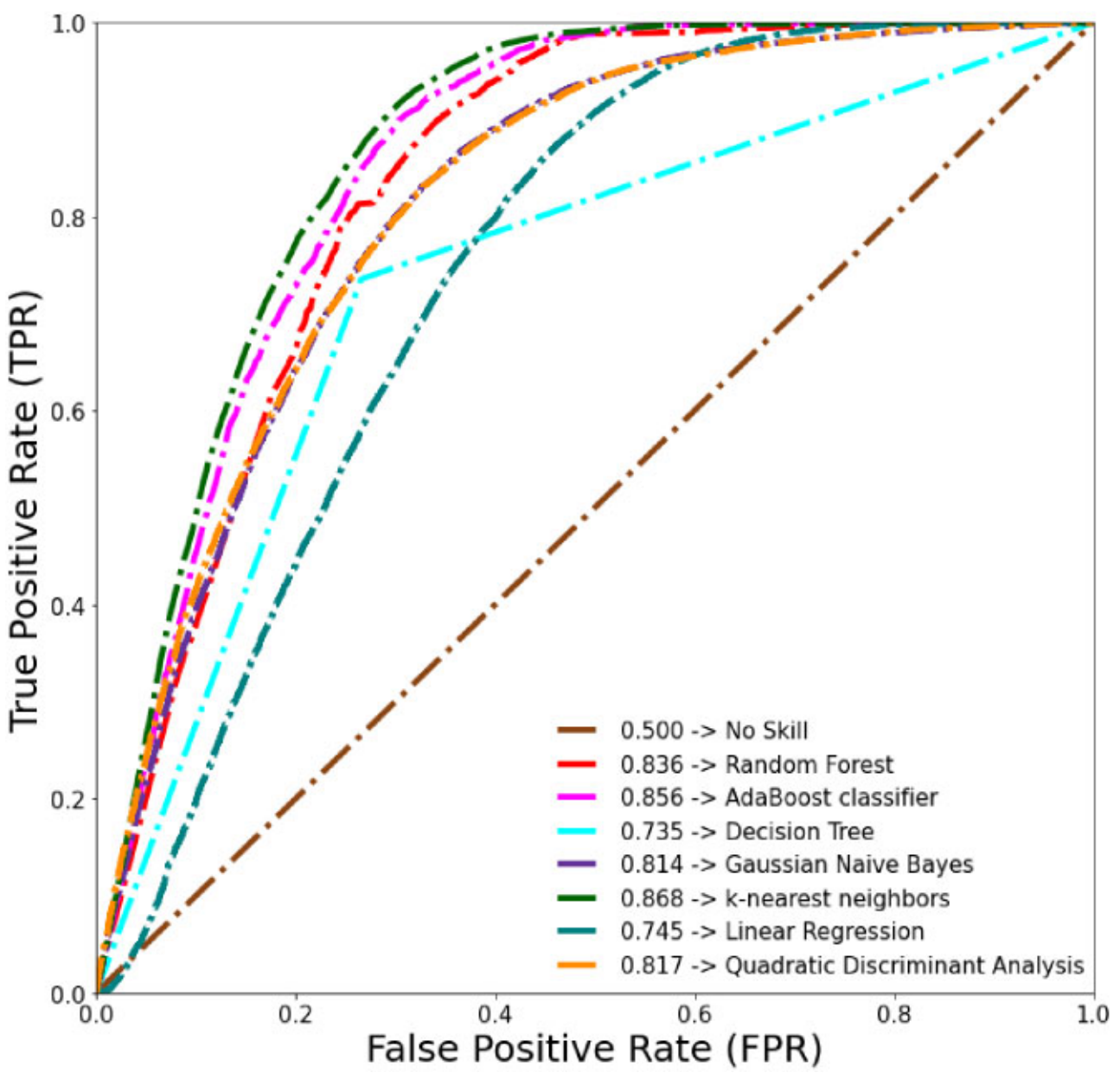

**Figure 3.** Plot of the ROC curve from seven different mechanisms of machine learning. This gives the relation between the rates given different thresholds for each algorithm. Since the bigger the area below the ROC curve, the more efficient the neural network, we can conclude which method was the best. Each one of them is identified in the legend. The brown dashed line is the guideline where the score of the machine learning is 0.500; all the curves above have a score bigger than the guideline. They gave us the following score: Random Forest: 0.836; AdaBoost classifier: 0.856; Linear Regression: 0.745; Gaussian Naive Bayes: 0.814; K-nearest neighbours: 0.868; Decision Tree: 0.735; Quadratic Discriminant Analysis: 0.817. Thus, we chose the K-nearest neighbours algorithm for our machine learning.

By looking at the results, we can see that for values of $\lambda$ above around 0.45, we have a significant number of galaxies which are photometrically similar to the VIPERS galaxies. Furthermore, there are some galaxies with such $\lambda$ values which are not classified as having a match.

### 4.2 K-dimensional tree

We could simply test our galaxy selection based on the value of $\lambda$ obtained in the previous subsection. However, we need to note that the VIPERS catalogue is not a fully complete selection of galaxies in the photometric domain. Usually, spectroscopic samples are sparse selections of the photometric counterpart. This means that within our analysis, there may be galaxies which simply do not have galaxies which are close to them in colour space, which would have passed the colour selections of VIPERs but were simply not chosen by a change of the spectroscopic selection. In order to do this, the aim of this subsection is to group galaxies within a *K-d Tree* structure and select voxels in colour space according to the average value of $\lambda$.

A K-d Tree (short for a K-dimensional Tree) is a data structure used for efficient search and retrieval of points in a K-dimensional space. It is a binary search tree where each node represents an axis-aligned hyperrectangle, also known as a bounding box, that partitions the space into two regions. The K-d Tree is constructed by recursively subdividing the space into two halves along the median of the data points on one of the K-dimensions at each level of thetree.

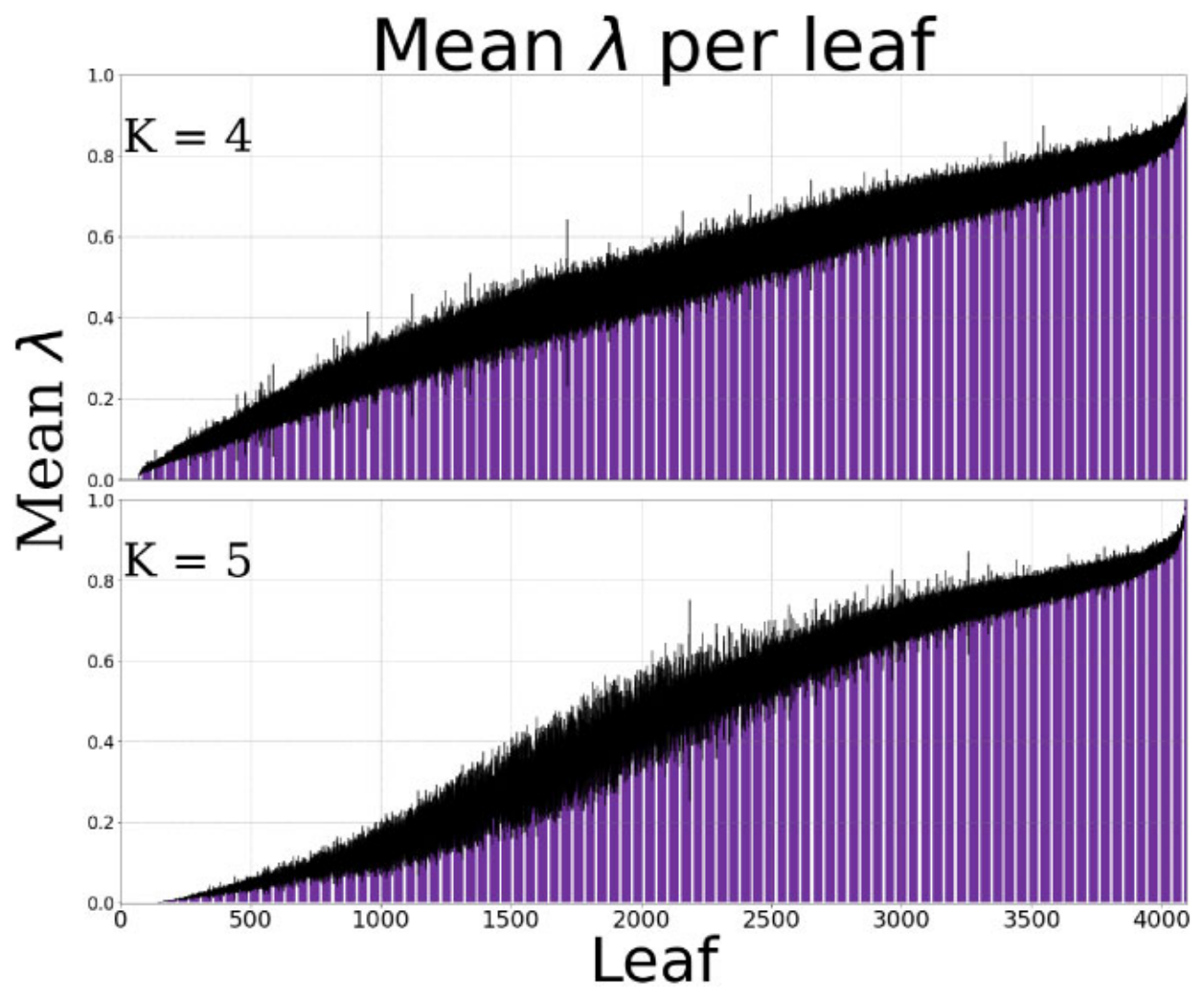

**Figure 4.** Mean λ on each leaf and its variance. We can see that the K-d Tree clearly classifies sections of colour space in a way that we can rank the likeness of the DES galaxies to the likeness of the VIPERS galaxies as a function of the leaf number. We obtain this with an associated error on the lambda and, therefore, can make a clear selection of the DES galaxies that most look like the galaxies contained within VIPERS in a simple, direct and unambiguous way.

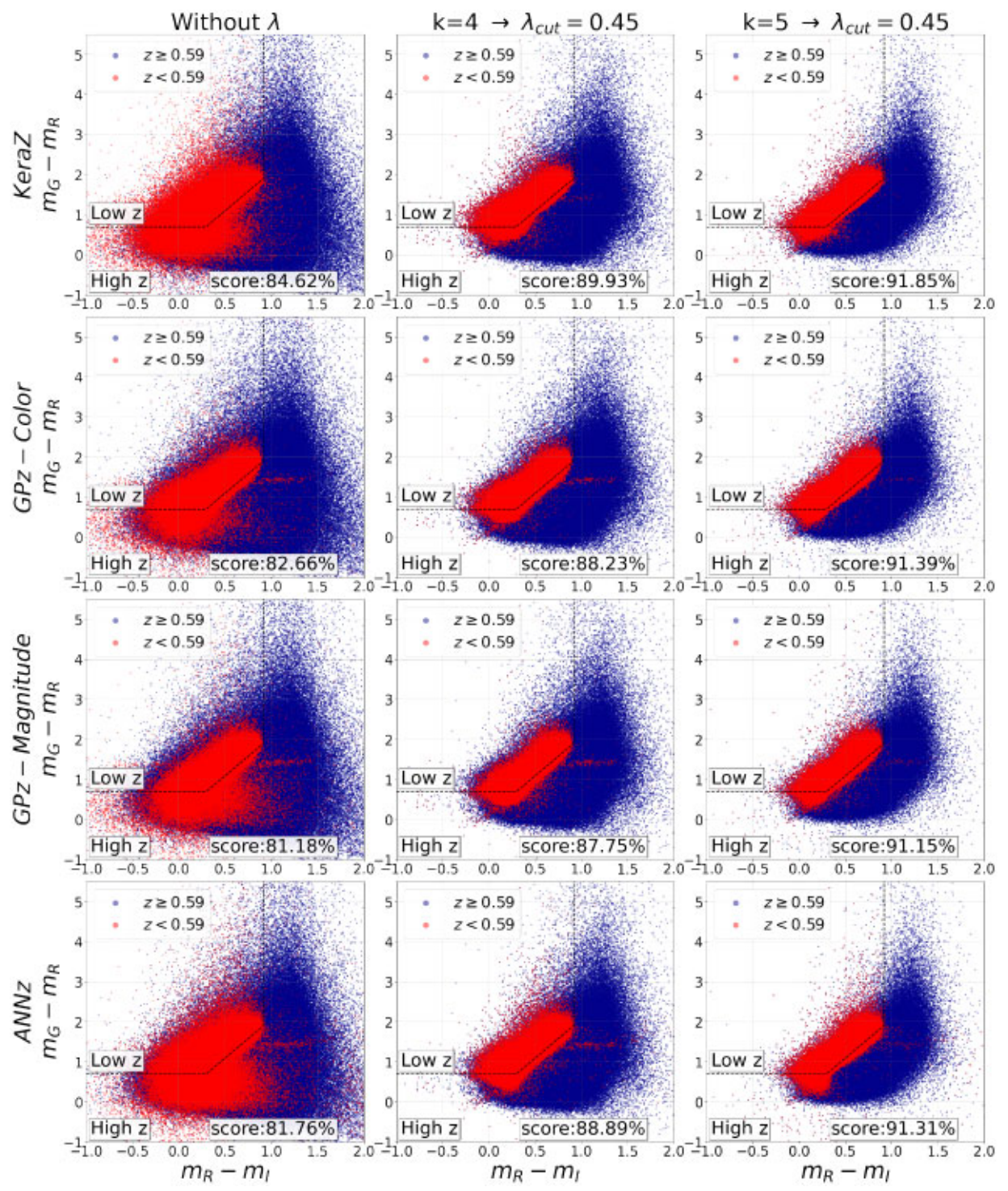

**Figure 5.** Comparison between the photometric redshifts in the colour–colour diagram with Lambda cuts. Galaxies were divided such that the orange points represent low-redshift galaxies while the blue points represent high-redshift ones, considering a best-fitting cut at $z_0 = 0.59$. With cuts at λ, i.e. disregarding galaxies with $\lambda < 0.45$, the classification shows higher accuracy, as can be clearly seen in the reduction of overlap between galaxies with lower redshifts (orange points) and higher redshifts (blue points) and the increasing in the classification score.

**Table 4.** Summary of galaxy counts and match statistics between DES and VIPERS catalogues. The fraction galaxies similar to VIPERS galaxies is defined as galaxies satisfying $\lambda > 0.45$. The matched DES-VIPERS galaxies are counted only in the area of the VIPERS survey. We can see that strictly speaking around 30% of the DES galaxies should have similar colours to the VIPERS galaxies, and that out final catalogue cut at 0.45 has just under 50% of the final catalogue, this means around 18% of the galaxies are slight extrapolations of the colours in terms of their photometric redshifts.

| Quantity | Value |
|---|---|
| Total DES galaxies ($N^{\rm gal}_{\rm DES,\,tot}$) | 247,191,790 |
| Total VIPERS galaxies ($N^{\rm gal}_{\rm VIPERS,\,tot}$) | 50 129 |
| Matched DES–VIPERS galaxy fraction within VIPERS | 0.1472 |
| Fraction of similar galaxies within VIPERS | 0.3019 |
| Fractions of DES galaxies surviving ($\lambda > 0.45$) | 0.4875 |

Once constructed, a K-d tree can be used to perform a variety of operations on the data, including *nearest neighbour search*, *range search*, and *K-nearest neighbour search*. Nearest neighbour search is the process of finding the closest point in the tree to a given query point. Range search is the process of finding all points within a certain distance of a given query point. K-nearest neighbour search is the process of finding the K closest points to a given query point.

In our work, we used two different configurations of K-d Tree. In the first one, we used four colours ($m_g - m_r$, $m_r - m_i$, $m_i - m_z$ and $m_z - m_Y$) as data, which implies that our K-d Tree has four dimensions (K = 4). The second one was adding to the data the magnitude observed by the filter I, thus K = 5.

On each leaf of this tree, we applied the machine learning methods used in the last section to obtain the lambda an estimate of the probabilities of DES galaxies being similar to VIPERS galaxies. As we can see in Fig. 4, the result is clearer and separated, which means that each method can better separate if a galaxy is photometrically similar or not. With this in mind, we were able to evaluate the mean value of λ on each leaf and its variance, as presented in Fig. 4.

Comparing both images in which the leaves were orderly from the lowest mean λ to the biggest one, we can see that when we added one more dimension (magnitude I) in the K-d Tree, the mean of λ of the leaves tend to lower its value. This implies that this Tree can determine more accurately that particular galaxies are similar to galaxies with VIPERS photometry, which means that it has a higher rate of TPR.

Each branch gives us the completeness λ for each galaxy, a variable from 0 to 1 that providing a parameter estimating how similar are the galaxies. As previously mentioned, the galaxy is not similar to many of the VIPERS galaxies when λ close to zero; thus, although we used a machine learning code to predict its redshift, we cannot use it as the reliability of its redshift will be low. While λ close to 1 indicates that the galaxy is similar to the ones we have the redshift, the probability of the machine learning code predicting $z$ that is close to a reliable estimate is high.

## 5 RESULTS

In this section, we present the results from our photometric redshift calculations together with the results from the calculation of the selection function chosen. We present the results in such a way as to maximize the quality of the photometric redshift evaluation while at the same time minimizing the error associated with the overall sample for the photometric redshifts. This is because we know that some of the samples will have poor redshift estimates, due to the incomplete coverage of DES with VIPERS, while other sections of the sample will have good estimates.

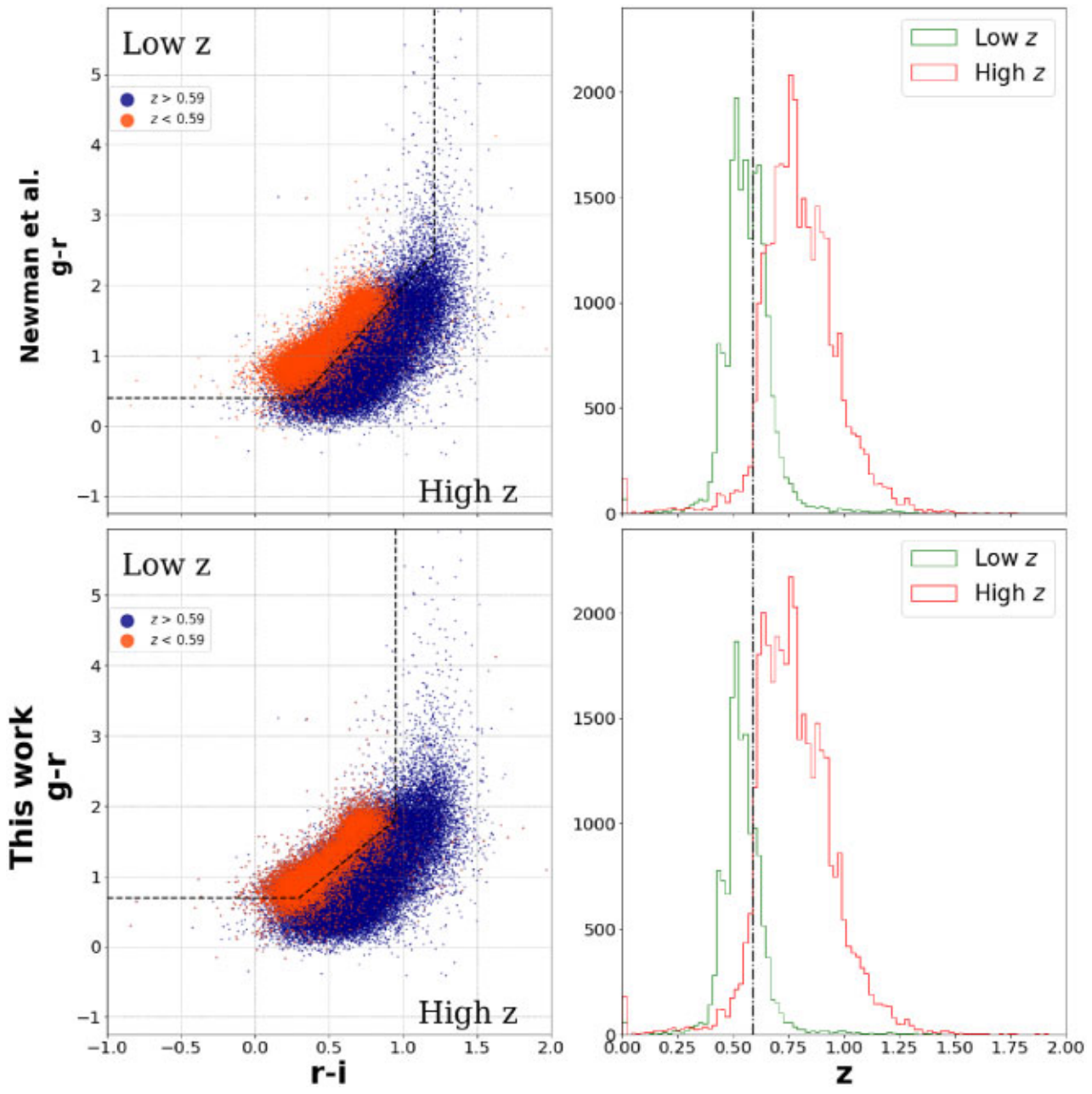

**Figure 6.** Comparison between spectroscopic redshifts in the colour–colour diagram ($m_g - m_r$ as a function of $m_r - m_i$). The scatter plot utilizes galaxies from VIPERS that are in the same sky region as DES. The spectroscopic redshift cut considered was $z_{\rm spec} = 0.59$, resulting in low redshifts with $z_{\rm spec} < 0.59$ and high redshifts with $z_{\rm spec} > 0.59$. The histogram, as a function of VIPERS spectroscopic redshift in the second column, displays galaxies with low redshifts in green and high redshifts in red.

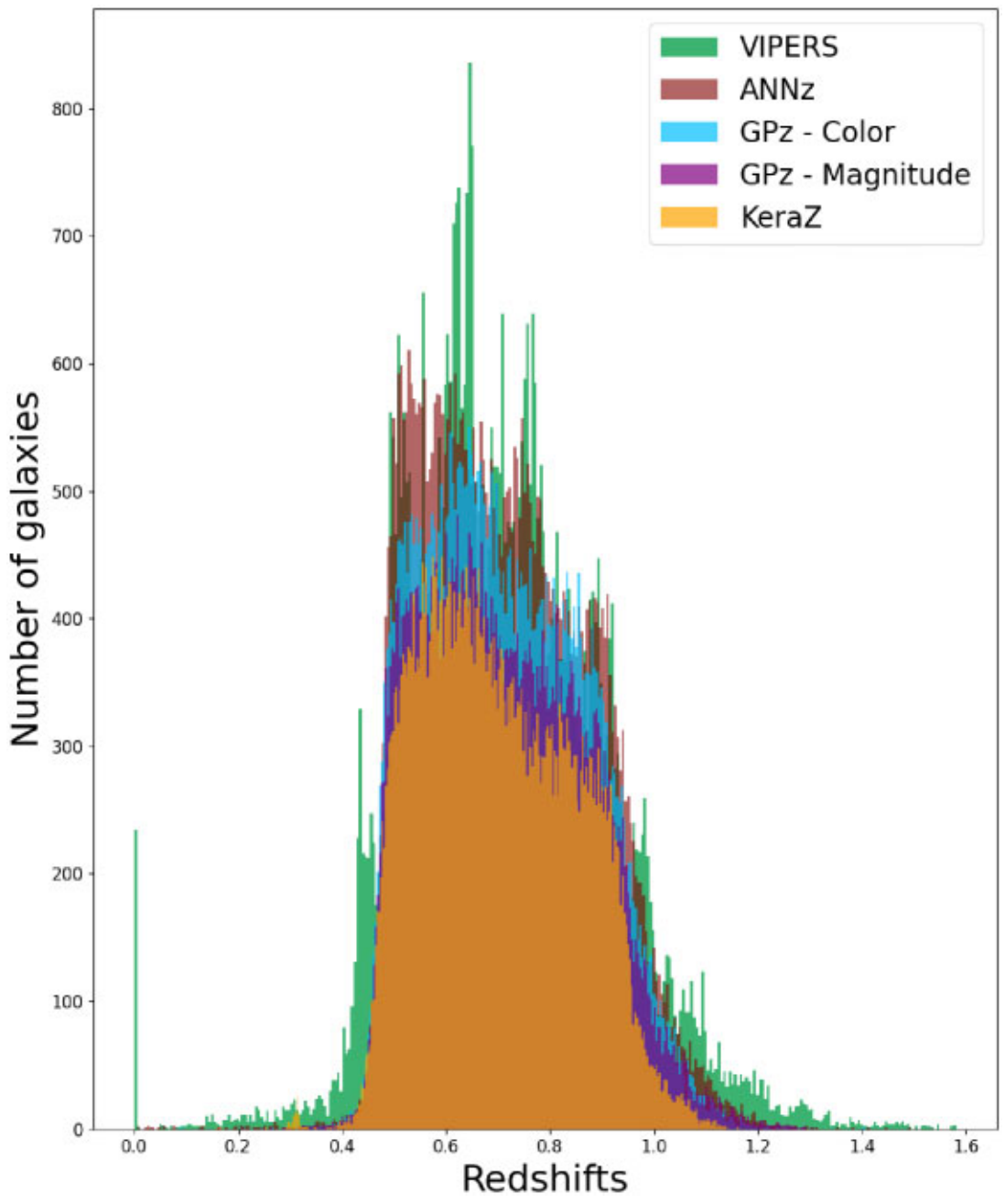

**Figure 7.** Histogram of the spectroscopic redshift and the photometric redshifts estimated by the machine learning algorithms presented on this paper. The green histogram represents the spectroscopic redshift from VIPERS. The maroon histogram shows the photometric redshift evaluated with the ANNz programme, while the orange histogram shows the photometric redshifts predicted by the KeraZ code. The blue histogram shows the photometric redshift evaluated through the GPz code using the four colours ($m_g - m_r$, $m_r - m_i$, $m_i - m_z$, and $m_z - m_Y$) and the purple histogram display the photometric redshifts predicted by the GPz code with the four colours and the magnitude in filter i. We plot this histogram from $z = 0$ until $z = 1.65$, there is a fraction of galaxies with higher redshift excluded from this plot. The percentages of galaxies excluded removed from the histograms are: GPz–Colour: 0.39 per cent, GPz–Magnitude: 0.16 per cent, KeraZ: 0.00 per cent, and ANNz: 0.02 per cent.

In previous sections, we presented each of the techniques used in this work and the procedure chosen to apply these same techniques. From now, we follow the same choices outlined in previous sections.

### 5.1 Reliability of the $z$ photometric

Before outlining the statistical properties of the photometric sample results, we start by treating the issue of homogeneity within the data set. We have studied this by using the calculated values for the $\lambda$ parameter, which were computed from previous sections. We effectively aim to determine to what extent a galaxy with a lower value of $\lambda$ has a less accurate photometric redshift than one with a large value of $\lambda$. In summary, the VIPERS data set is not a complete and reliable representation of the DES data. We can only obtain reliable photometric redshifts from galaxies that can reliably be compared to the training set from VIPERS.

To perform this study, we use a selection mechanism which was employed in the DEEP2 galaxy survey (Newman et al. 2013) aiming to select high redshift galaxies with optical data and attempt to reproduce it with our photometric redshifts. The DEEP2 selection was based on a series of BRI colour cuts, which are outlined in Newman et al. (2013). These colour cuts are designed to very accurately separate, in this two-dimensional colour space, galaxies above and below redshifts of the order of 0.7. Comparing some fields where these cuts were implemented, and one test field where the cut was not implemented, the DEEP2 analysis has shown that this selection effect is very effective at finding high redshift galaxies and also at selecting galaxies above and below $z \sim 0.7$.

Similar selections have been inspired by this analysis and have performed target selection in a two-dimensions colour plane. For instance, a *grz* selection has been applied to produce a target data set for the eBOSS sample (Raichoor et al. 2017), where the colour cut has been designed to optimally select galaxies between $0.7 < z < 1.1$. Obviously, the BRI filters do not exist in the DES data set; however, they are not too dissimilar from the *gri* filters within DES, B being very similar in terms of wavelength from the $g$ filter, $R$ is similar to the $r$ filter, but wider and the $I$ and $i$ filter again are also not too dissimilar. Therefore, a cut similar to the DEEP filter in the *gri* plane would most likely separate galaxies above and below a given redshift $z_0$ to be determined, which should not be dissimilar from 0.7 but is expected to be somewhat different given the differences in filters. The DEEP2 colour cuts were essentially two colour cuts in the $m_B - m_R > 0.389$, $m_R - m_I < 1.211$, and $m_B - m_R > 2.45(m_R - m_I) - 0.311$.

We find that if we simply apply these cuts, ignoring the difference in filters to the DES sample, indeed, we have a very clear separation of spectroscopic galaxies at the redshift $z_0 \sim 0.59$, by applying the same cuts to the VIPERS testing sample, using $m_g - m_r > 0.389$, $m_r - m_i < 1.211$, and $m_g - m_r > 2.45(m_r - m_i) - 0.311$. However, we also find that these limits can be optimized if we use slightly different boundaries of the selection criteria of $m_g - m_r > 0.691$, $m_r - m_i < 0.95$, and $m_g - m_r > 1.90(m_r - m_i) - 0.298$. These will be our modified DEEP2 cuts to select high redshift galaxies in this section.

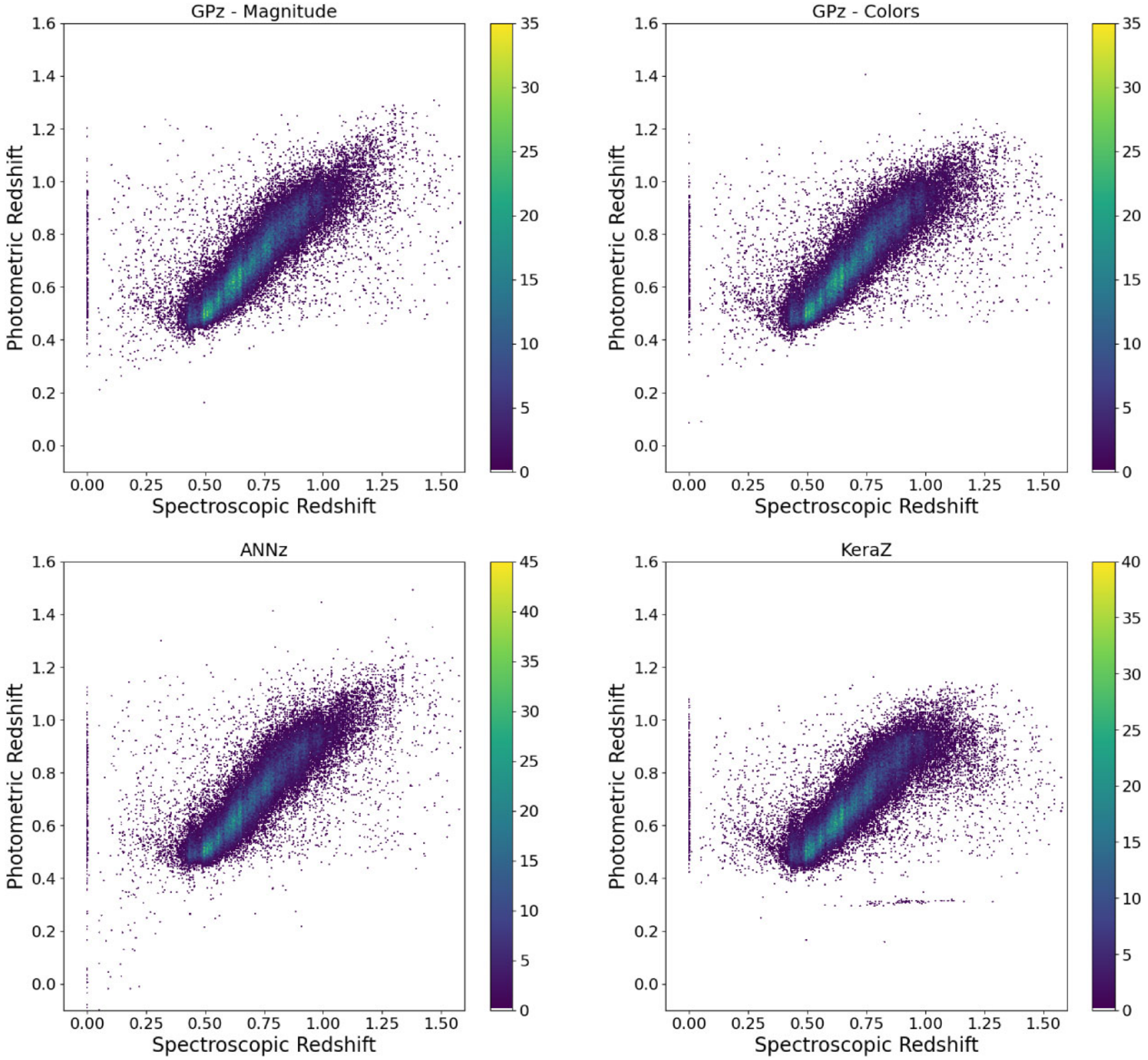

**Figure 8.** Scatter plot of the photometric redshift of each machine learning as a function of the spectroscopic redshift. We can see how similar the results are. It is possible to observe some differences; for instance, in the photometric redshift calculated by KeraZ, we have some concentration of galaxies with $z_{\rm phot} \sim 0.3$ regardless, which is likely due to the effects of training arising from the large-scale structure of the Universe present in the training data. We also see that, for redshifts greater than 1, the calculated values of GPz Colour and KeraZ are a bit further away from the diagonal. However, overall, the statistics of these curves are very similar. The colourbar represents a relative density of the sample.

After running the K-d Tree algorithm to perform a calculation of $\lambda$, we can now organize the photometric galaxies within DES in ascending order accordingly with the mean $\lambda$ of the galaxies in each branch. From this, we can plot the mean $\lambda$ from each K-cell within the K-d Tree with its respective variance from both K-d Trees. After that, we use this information together with the photometric redshift to be able to make cuts in $\lambda$, which will indicate a photo-$z$ reliability of the selected galaxies.

To contextualize the reliability of the photometric redshift estimates and the selection strategy employed, Table 4 summarizes key global statistics of the matched DES–VIPERS sample. The total number of galaxies in the DES DR2 data set exceeds 247 million, while the VIPERS sample used for training and validation comprises just over 50 000 galaxies. Among these, approximately 14.7 per cent of DES galaxies have counterparts in the VIPERS footprint, and only 30.19 per cent are considered 'similar' according to the $\lambda$-based selection. By applying a threshold cut of $\lambda > 0.45$, we identify a high-confidence subset of galaxies for which the photometric redshift estimates are considered reliable because their colours are similar to VIPERS colours. This final catalogue is a subset of the total DES galaxies and accounts for approximately 48.75 per cent of the full catalogue.

We plot our findings in this section in Figs 5 and 6. In Fig. 5, we show the effect of a $\lambda$ cut for the galaxies shown in the gri colour–colour plane, with and without a lambda cut of 0.45. We perform the analysis with colours only (K = 4) and with colours and magnitudes (K = 5). The galaxies plotted are the photometric galaxies, and the colour indicates if those galaxies have a photometric redshift above (blue) or below (orange) 0.59. We can see that the lambda cut works, and in all cases, to help constrain the galaxies at a low photometric redshift at values within the colour bounds that we have defined above. This is exactly what is to be expected if the photometric redshifts of the galaxies with a lambda below 0.45 are less reliable, therefore being 'misclassified' in terms of redshift.

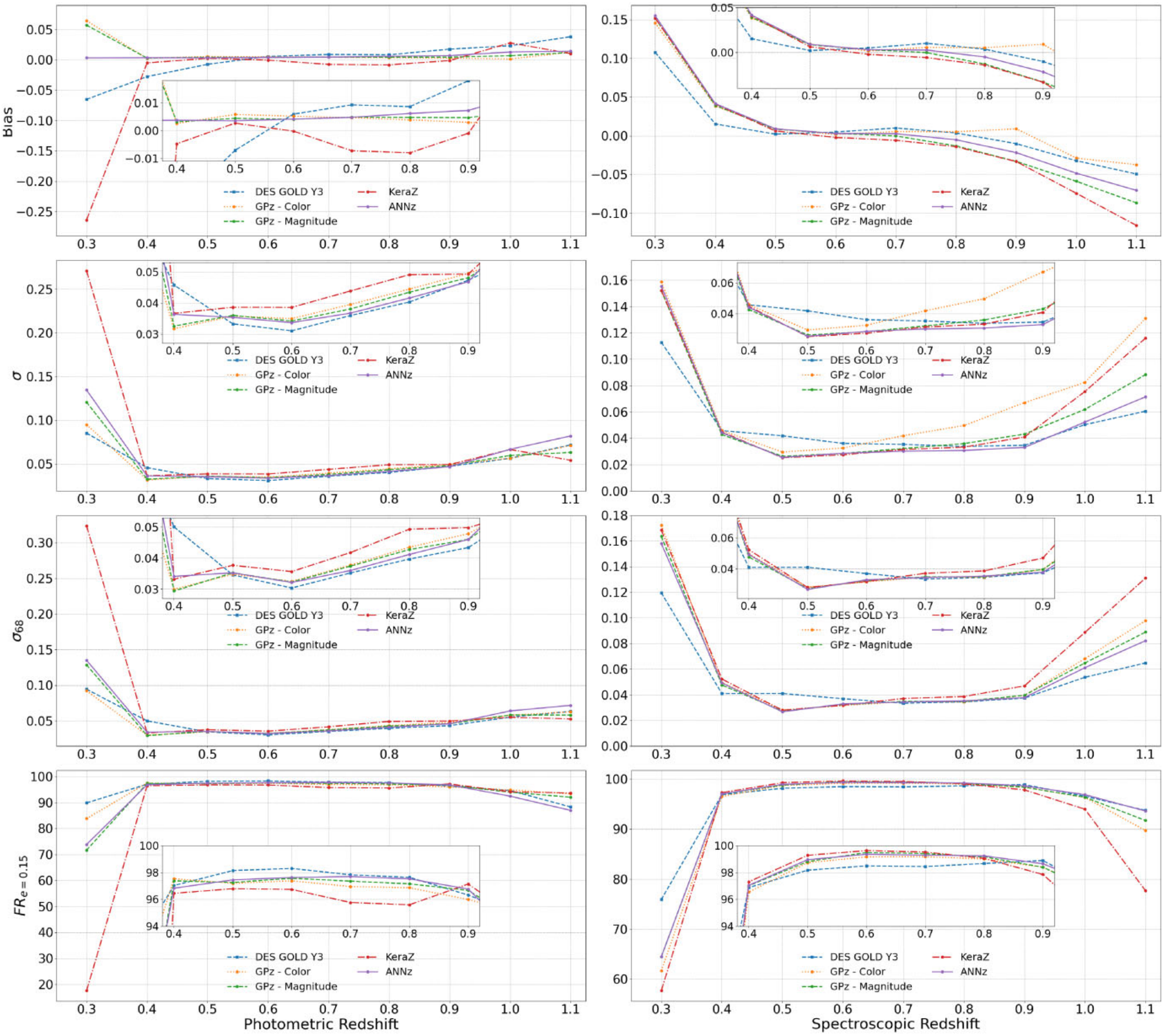

**Figure 9.** Plots of the metrics detailed in Section 3.4 for redshift bins with length 0.1. Each metric was evaluated with the photometric redshift of the four machines learning methods used in this work and assuming the spectroscopic redshift from VIPERS. The bins start at 0.3 and go up to 1.1. The yellow curves represent the metrics calculated with the photometric redshift evaluated through the GPz code with four colours ($m_g - m_r$, $m_r - m_i$, $m_i - m_z$, and $m_z - m_Y$) while the green curve shows the metrics evaluated with the photometric redshift predicted by the GPz code that uses the four colours and the magnitude in filter i. The red curves represent the metrics calculated with the photometric redshift evaluated through the KeraZ code, while the purple curves show the metrics evaluated with the photometric redshift predicted by the ANNz2 code. Each line displays on the corner left the type of metric used on each row. All these galaxies are part of the data set with the match of the VIPER's galaxies and DES's DR2 galaxies. In the first column, the galaxies were separated through the value of the photometric redshift, while in the second column, they were separated in bins based on the value of their spectroscopic redshift. To confirm the quality of our results, the blue curves represent the metrics applied to a DES Y3 GOLD sample with the same galaxies as our work.

This is specifically to be expected at the low redshifts, especially given that the DES galaxies have not got any information from the *u* band from these galaxies, being able to identify the 4000 Angstrom break all the way to zero redshifts. Therefore, the compactness of the scatter plots within the region defined by our cuts indicates that the lambda cut is indeed likely removing galaxies for which we have poor redshift knowledge at the low redshifts.

Nevertheless, this is not the only feature seen. We can see that the lambda cut is also removing a large fraction of blue objects at the low $m_g - m_r \sim 0$ and $m_r - m_i \sim 2.0$ region. These are likely to be higher redshift galaxies, which are simply not constrained within the VIPERS data set, probably because of the limited wavelength reach of the instrument used for this survey. It is clear that galaxies above a redshift of around 1.3 are missing from the VIPERS catalogue as they fall within the redshift desert, and IR detectors are needed to accurately find their spectroscopic redshift.

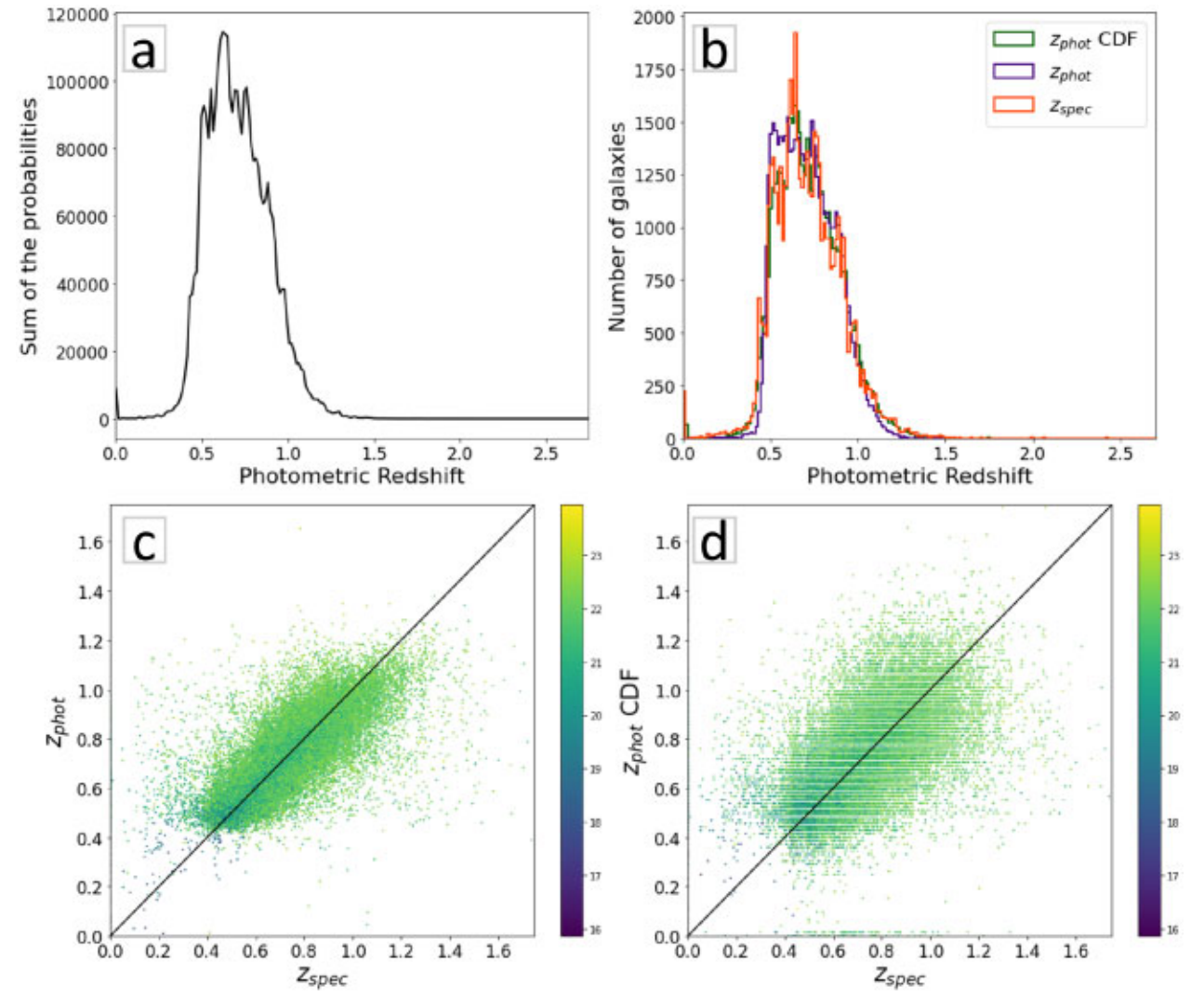

**Figure 10.** Plots of the ANNz2 photometric redshifts with the use of the probability function distribution (PDF) given by the code. The image a shows the histogram of the sum of all 200 PDFs from all galaxies in the data set as a function of the redshift. The image b displays the histogram of the photometric redshift, the spectroscopic redshift, and the photometric redshift evaluated through the use of cumulative distribution function (CDF). The image c is a scatter plot of the photometric redshift from the ANNz2 as a function of the spectroscopic redshift while the image d is a scatter plot of the photometric redshift evaluated with the CDF as a function of the spectroscopic redshift.

The final feature that is present in this Fig. 6 that we comment upon are some faint horizontal branches that exist in the photometric redshifts with values below 0.59. A horizontal branch appears in this sample encroaching in the photometric region, which is expected to have high redshifts. We have not investigated this feature in-depth, but this is possibly the result of a small amount of star contamination in this sample. As it can be seen from the DEEP2 sample selection plots in (Newman et al. 2013), there is a track of stars that cuts the low redshift region and extends into the high redshift region in this plot. Any of these residual stellar contaminations could appear in a photometric redshift analysis as having a preferred lower redshift, as the overall colours would be identified as being consistent with a redshift of zero.

We see in Fig. 8 the same scatter plots for the case of the spectroscopic redshifts contained within the VIPERS sample, again colour-coded in terms of their redshifts being above or below 0.59. We can see that. Indeed, the modified DEEP2 colour cuts are extremely effective at finding galaxies at the highest and lower redshift ends of VIPERS, and this indicates the reliability of the lambda cut of around 0.45, which we made for the photometric redshift sample.

### 5.2 Quality of the photometric redshifts for the DES DR2:VIPERS sample

We have applied the KeraZ, ANNz2, and GPz algorithms on the DES DR2 data set as described in this paper. We present in Figs 7–8 the summary of our findings.

These data were obtained from the merge between data from VIPERS PDR-2 with DES DR2 (hereinafter DR2 MAIN). The photometric analysis in DR2 MAIN used the *MAG_AUTO_DERED* (i.e. the DES model magnitude correction for dust) as galaxy magnitudes and the *MAG_AUTOERR* as magnitude errors. The training and validation set were performed using the colours $m_g - m_r$, $m_r - m_i$, $m_i - m_z$, $m_z - m_Y$, and the $i$-magnitude. The results are presented for colour-selected and magnitude-selected training of the DR2 MAIN test data. As we can see from Fig. 7, the histograms of the spectroscopic redshifts and photometric redshifts are similar for all methods, and there are some differences, mainly the fact that the spectroscopic sample has shot noise. This shot noise comes from the large-scale structure in our Universe but also that usually, within a given redshift range, photometric redshifts are biased high on the low redshift end and biased high on the high redshift end. Which can be seen from these histograms.

We can clearly visually see from Fig. 8 that the quality and level of the outlines and the scatter of these photometric redshifts applied to the testing set are high. The statistics are provided in Fig. 9, and we can confirm that the scatter of 68percnt of the sample as a function of redshift is of the order of 0.035, which is of excellent quality. The level of outliers is placed at the per cent level. and the quality of these redshifts in the range from 0.5 to 0.9 is very high compared to other available photometric redshifts in the literature. We do not know that the bias as a function of the photometric redshift is slightly larger in the KeraZ method, which is a similar implementation to the first version of ANNz2. This was found to be due to the high flexibility of the neural network when many intermediate layers are present and is a feature that does not exist in the second version of ANNz given many other training methods were used in the code.

We also compared our results to a subset of the DES Y3 GOLD sample which was matched to VIPERS confirm our findings, as described in Sevilla-Noarbe et al. (2021), shown in Fig. 9. We found that our implementations exhibit a significant lower bias in the redshift range $0.4 \leq z < 0.6$ while maintaining overall consistency with the DES Y3 GOLD performance across the full redshift range. In terms of the dispersion metrics ($\sigma$ and $\sigma_{68}$), we find that GPz Colour and GPz Magnitude approaches provide competitive and slightly better results compared to DES Y3 GOLD, especially in the intermediate redshift range ($0.4 \leq z \leq 0.6$), where they exhibit similar or slightly lower values of $\sigma$ and $\sigma_{68}$. In the high-redshift regime ($z \geq 0.9$), GPz–Magnitude shows improved stability. Finally, for the $FR_{e=0.15}$, in the intermediate redshift range ($0.4 \leq z \leq 0.9$) GPz-based approaches, exhibit comparable performance to DES Y3 GOLD sample, although the DES GOLD sample is of better quality. At high redshifts ($z \geq 0.9$), all methods experience a mild increase in $FR_{e=0.15}$, but GPz–Magnitude appears to be the most stable alternative. We however note that the sample from DES does not provide which galaxies are good matches to VIPERS, which our data set does; furthermore, the overall dispersion of our results can be compared to the subsample from Sevilla-Noarbe et al. (2021) and also compared to the dispersion of the full DES gold sample by looking at fig. 11 of Sevilla-Noarbe et al. (2021). We can conclude that the subsample of VIPERS-like galaxies from DES has a better dispersion and better photometric redshifts than the full DES gold sample, and that our results improve slightly on the Sevilla-Noarbe et al. (2021) results. Therefore, we conclude that this analysis can subdivide the sample in such a way that it could be more suited for large-scale structure studies.

Furthermore, we also implement the redshift estimation within ANNz2 in a PDF format, where instead of the output being the value of the redshift, the output consists of a large number of outputs representing the probability of the galaxy being within a given part of the redshift distribution. These results are shown in Fig. 10, where we show the ANNz2 results using a PDF output, including the most probable redshifts as well as a CDF redshift, which is a redshift drawn randomly from the PDF output for each galaxy. These results remarkably show that the histogram of the CDF photometric redshifts from ANNz2 match extremely well with the spectroscopic redshift of the testing set. What it could be expected as the redshift distribution of the photometric redshifts, in this case, is learning directly from the spectroscopic redshift $n(z)$. This is seen from the upper right panel of Fig. 10. The CDF estimation method has slight differences compared to the AvgPDF-ANNz2, KeraZ, and GPz methods. The value of the photometric redshift used to compute the metrics for each estimator is $0.1 < z < 1.2$, and if compared with the CDFs estimators with the other, we can see that is an excellent agreement.

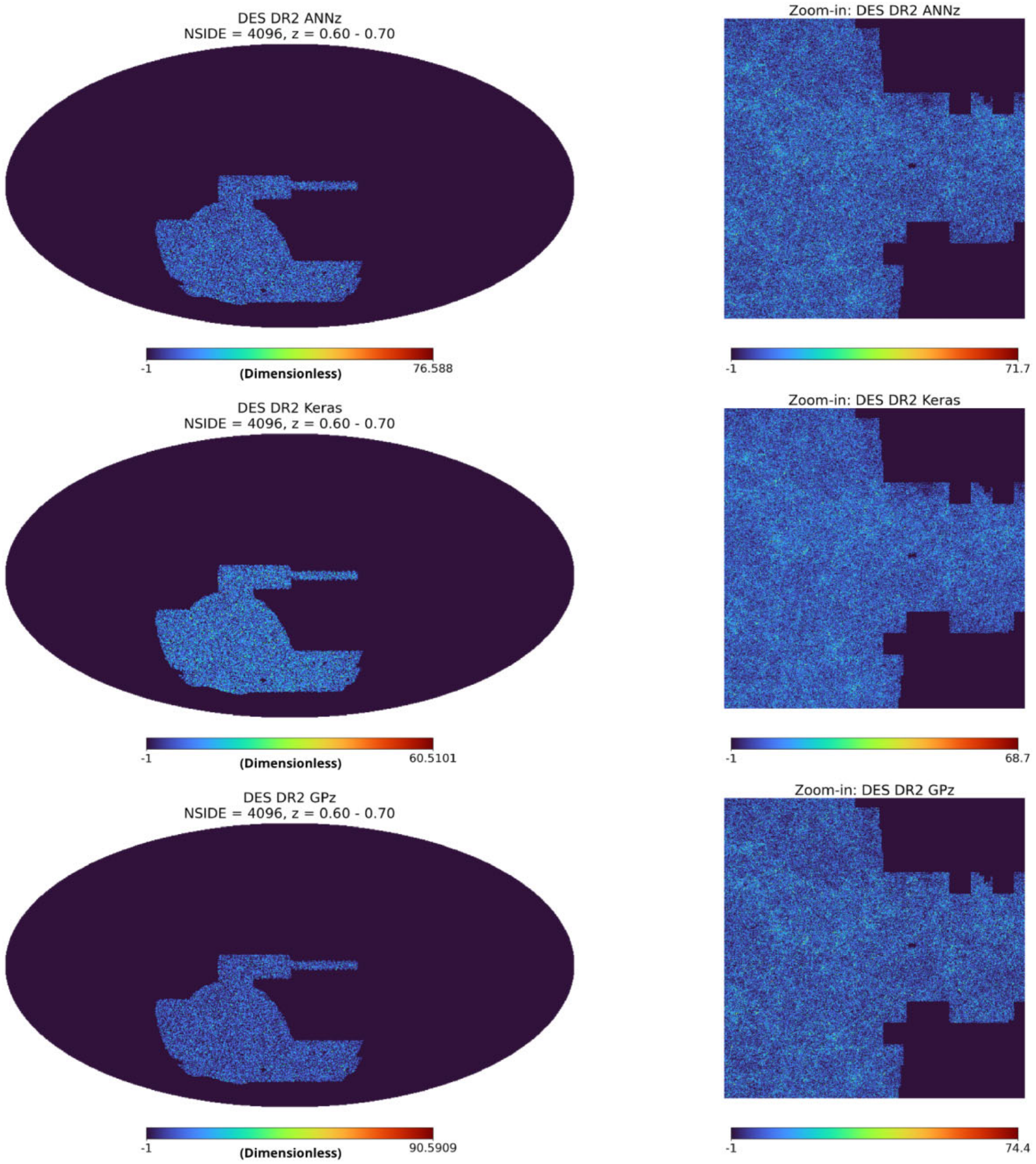

**Figure 11.** Results obtained for the use of algorithms on DES release 2 data. On the left panels, we show the overdensity maps for each algorithm in the same redshift bin, defined as $\delta = (n - \bar{n})/\bar{n}$. On the right panels, we present zoom-ins of the same maps, highlighting finer details of the data coverage. All maps are dimensionless, and the colour bars represent the relative density of objects at each point compared to the mean. The highlighted regions emphasize the ability of each method to recover the survey footprint and local density contrasts.

Finally, we present in Fig. 11 the maps of these photometric redshift catalogues. We can see that they seem to be tracing the large-scale structure of our Universe, and these should be maps that can be used for the study of cosmology via their statistics.

## 6 CONCLUSIONS

We have in this paper performed an analysis of the photometric redshifts of the publicly available DES. We decided to concentrate on a single training data set to be able to extract redshift data from the photometric catalogue. We used a sub-sample of the training set methods that exist in the literature, namely GPz, ANNz2, and KeraZ. ANNz2 has a host of training methods within its subroutines, and the KeraZ, which is based on Keras, is a publicly available neural network implementation of deep learning networks.

We described the data sets used and outlined a method in order to attempt to overcome the fact that no current training set data is completely matched to a photometric survey complete such as DES. Aiming to do this, we have run training algorithms which would be able to identify how likely a given galaxy from DES would be well-matched in terms of colour with galaxies from the training set that we have available from VIPERS. We performed some tests based on well-established colour cuts in the literature in order to assess the reliability of these photometric redshifts, and we conclude that a cut of around 0.45 in this likeness parameter $\lambda$ suggests that the galaxies have reliable photometric redshifts for values larger than $\lambda \sim 0.45$. We recommend the use of this cut for this sample.

We then present the results for our photometric redshift calculations and show an excellent recovery of the spectroscopic redshifts of the testing set, which indicates that the photometric redshift sample should be of excellent quality for cosmological studies. We present the levels of scatter and bias present within the sample for many methods and conclude that the methods used are robust, though there is some internal difference between them. This further indicates that the quality of the redshifts is high. Finally, we show some preliminary cosmological maps for this sample, which we plan to use in future analysis for cosmological studies.

## ACKNOWLEDGEMENTS

FBA acknowledges this work has been supported by a USP visitor grant as well as a USTC start-up grant. ARQ acknowledges FAPESQ-PB support and CNPq support under process number 310533/2022-8.

## DATA AVAILABILITY

The data underlying this article can be shared on a reasonable request by the corresponding author. They will be freely accessible after the publication of all projects planned by our group from the repository https://github.com/zxcorr/PhotoZ-DES.

This paper has been typeset from a TEX/LATEX file prepared by the author.